\documentclass[manuscript,preprintnumbers,amsmath,amssymb,floatfix]{revtex4}

\usepackage{graphicx}
\usepackage{dcolumn}
\usepackage{bm}
\usepackage{amsmath}
\graphicspath{{./},{./Figure/}} 

\begin{document}


\title{Analytical model of atomic-force-microscopy force curves in viscoelastic materials exhibiting power law relaxation}

\author{J. S. de Sousa, J. A. C. Santos, E. B. Barros } 
\affiliation{Departamento de F\'isica, Universidade Federal do Cear\'a, Caixa Postal 6030, 60440-900, Fortaleza, Cear\'a, Brazil}



\author{L. M. R. Alencar}
\affiliation{Instituto Federal de Educa\c{c}\~ao, Ci\^encia e Tecnologia do Cear\'a, R. Engenheiro Jo\~ao Alfredo s/n, 61600-050 Caucaia, Cear\'a, Brazil}

\author{W. T. Cruz, M. V. Ramos}
\affiliation{Departamento de Bioqu\'imica e Biologia Molecular, Universidade Federal do Cear\'a, Caixa Postal 6033, 60440-900, Fortaleza, Cear\'a, Brazil}

\author{J. Mendes Filho}
\affiliation{Departamento de F\'isica, Universidade Federal do Cear\'a, Caixa Postal 6030, 60440-900, Fortaleza, Cear\'a, Brazil}



\begin{abstract}
We propose an analytical model for the force-indentation relationship in viscoelastic materials exhibiting a power law relaxation described by an exponent $n$, where $n=1$ represents the standard viscoelastic solid (SLS) model, and $n<1$ represents a fractional SLS model. To validate the model, we perform nanoindentation measurements of poylacrylamide gels with atomic force microscopy (AFM) force curves. We found exponents $n<1$ that depends on the bysacrylamide concentration. We also demonstrate that the fitting of AFM force curves for varying load speeds can reproduce
the dynamic viscoelastic properties of those gels measured with dynamic force modulation methods. 
\end{abstract}

\maketitle

\section{Introduction}

Nanoindentation methods have become  extremely important in the last
couple of decades due to the increasing interest in the study of the
viscoelastic properties of nano-systems. Viscoelasticity is typical of a number of materials: polymers, plastics,
composites, metals and alloys, building materials, and
biological tissues. Studying viscoelasticity in micro and nano scale
is crucial to shed new light onto the understanding of a wide range
of practical problems like drug delivery by nanoparticle carriers
\cite{fiel2011}, biomechanics of living cells and their response to
external forces \cite{hoffman2009,bremmel2010,rebelo2014}, and the possibility
of diagnosing diseases at early stages
\cite{darling2006,darling2007,ketene2011, fritsch2010, rebelo2013}.

Nanoindentation with the atomic force microscope (AFM) is one the
most popular methods to probe soft samples. Conventional AFM force
curves can be fitted with an appropriate model to
extract the materials properties, whereas the models from Hertz
\cite{hertz,sneddon1965} and Oliver-Pharr
\cite{oliverpharr1992,oliverpharr2004} are the most used ones. The
former is mainly used to study thin soft films and biological
samples (tissues and cells). The latter is usually applied to hard
materials like metals, glass and plastic. Both models consider that
samples can be described by a purely elastic solid. On the other
hand, viscoelastic materials are usually described in terms of the generalized Maxwell model, which is composed of associations of springs and dashpots, where each
spring element has a spring constant $E_i$, and each dashpot has a
relaxation time $\tau_i$ \cite{fung}. This model comprises four
particular cases: the elastic case (represented a single spring
element), Kelvin-Voigt model (represented by a spring and dashpot
connected in parallel), Maxwell model (represented by a spring and
dashpot connected in series) and the standard linear solid (SLS)
model (represented by the combination of a spring element connected
in parallel with a Maxwell arrangement of spring and dashpot).
The Maxwell model does not describe creep or recovery, and the
Kelvin-Voigt model does not describe stress relaxation. The SLS model is
the simplest model that predicts both phenomena, but it fails to
describe materials with more than one relaxation time scale. One can progressively add more Maxwell elements in order to describe materials with multiple relaxation times. The aforementioned models are able to describe materials exhibiting exponential shear relaxation in time $G(t) \propto e^{-t/\tau}$. 

However, there are many classes of materials (e.g. living cells, hydrogels, cross-linked polymers, colloidal suspensions and foams) whose viscoelastic properties are described by a power-law relaxation $G(t)\propto t^{-n}$, which cannot be modelled by an association of spring and dashpots (unless a very large number of elements is added) \cite{fabry2001,djordjevic2003,jaishankar2012}. The natural framework to model power law rheology is the fractional calculus \cite{podlubny1999}. Analogously to the viscoelastic elements (springs and dashpots), one defines a fractional element (Scott-Blair element) whose constitutive stress-strain equation is $\sigma(t) \propto d^n \epsilon (t)/ d t^n$, where $0 \le n \le 1$ and $d^n / d t^n$ is the fractional derivative operator. The fractional element interpolates between two responses: for $n = 0$ ($n=1$) one obtains the Hookean elastic spring (Newtonian dashpot) constitutive equation \cite{scottblair1947, schiessel1995, jaishankar2012}.

Several methods have been proposed to obtain viscoelastic properties of samples with the AFM. Darling \textit{et al.} modified the closed-loop feedback control of the z-axis movement to perform stress relaxation tests in their samples \cite{darling2007,ketene2011}. This approach allows the determination of intrinsic relaxation times of viscoelastic materials. Some groups modified the AFM to perform force-modulated dynamic rheology \cite{rebelo2014, fabry2001,mahaffy2004,alcaraz2002,nalam2015}. This allows the determination of the exponents of the power law response of viscoelastic materials in the frequency domain $G(\omega) \propto \omega^n$, but demands complicated modifications in the AFM apparatus as well. Within the framework of conventional AFM force curves, most of the studies in the literature only address the instantaneous elasticity modulus of the materials, disregarding viscoelastic effects because of the lack of simple models to extract the viscoelastic properties from the force curves. Ren \textit{et al.} measured the frequency-dependent instantaneous elasticity moduli $E_0$ of cancer cells subjected to anticancer drugs by changing the loading speeds $v_L$ \cite{ren2015}. Although their forces curves clearly show viscoelastic effects, they focused on the analysis of the power law relationship $E_0 \propto f_z^m$ ($v_L \propto f_z$) to classify the action mechanisms of the anticancer drugs. Some of the authors made use of an empirical model to determine the apparent viscosity of living cells and asphalt binder directly from force curves \cite{rebelo2013, rebelo2014b}. The major limitation in the investigation of viscolelastic effects by means of force curves is the loading speed of the cantilever. High loading frequencies (typically above 10-30 Hz) induces cantilever oscillations that reduces the accuracy of the curve fitting \cite{ren2015}. In this regard, Chyasnavichyus \textit{et al.} proposed the application of the known frequency-temperature superposition to surpass the limited range of AFM loading rates, and constructed the master curves of PnBMA polymers by fitting AFM forces curves measured at different temperature and loading rates with a viscoelastic force model based on the SLS model \cite{tsukruk2014}. None of the aforementioned works considered power law relaxation mechanisms in the viscoelastic modeling of the materials. 
 
In this work, we present an analytical model for the AFM force curves of viscoelastic samples indented by axisymmetric indenters that accounts for power law viscoelastic relaxation. We validate our model numerically with Finite Elements Modeling using a computational two-body indentation model described elsewhere (see Supplementary Material) \cite{santos2012}, and apply it in the AFM study of polyacrylamide gels. Polycrylamide gels are often considered as standard to test viscoelastic models of soft samples, with many works reporting their properties \cite{nalam2015,gavara2010,abidine2015}. We demonstrate that the measured properties of the gels are in good agreement with other studies in the literature, and that our model is able to extract viscoelastic properties that could only be previously determined with dynamic rheology methods. 

\begin{figure*}[ht]
\center \includegraphics[scale=0.7]{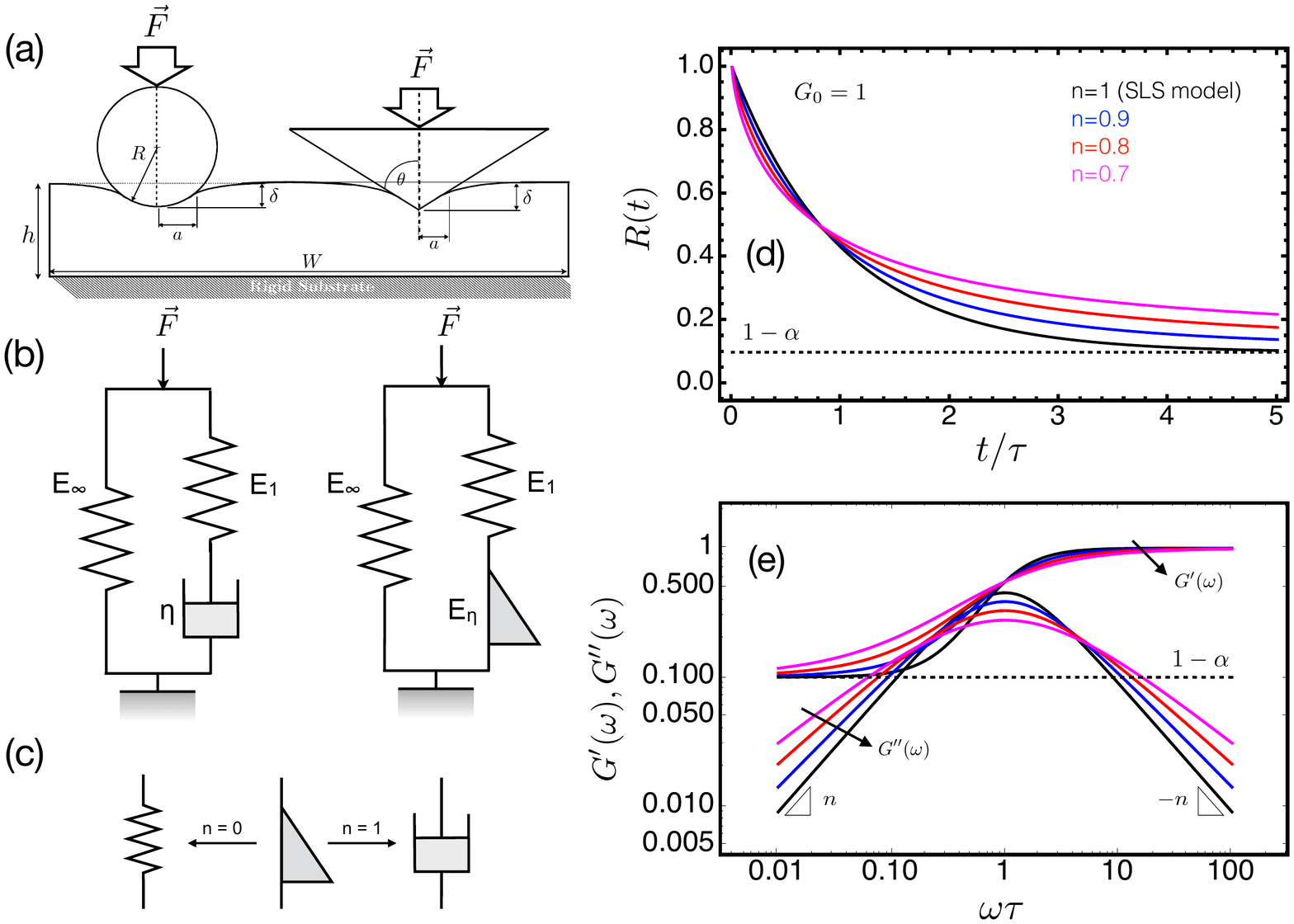}
\caption{(a) Schematics of the indentation of soft samples by conical and spherical indenters. The sample thickness $h$ is much larger than the maximum indentation $\delta_{max}$ in order to avoid finite thickness effects. (b) Schematics of the regular (left) and fractional (right) SLS model where the dashpot is replaced by a fractional element \cite{scottblair1947, schiessel1995, jaishankar2012}. (c) Representation of the fractional element that interpolates between Hookean spring and Newtonian dashpot. (d) Relaxation function of the fractional-like SLS model given by Eq. \ref{eq:SLSrelax}. (e) Frequency dependence of the storage and shear moduli given by Eq. \ref{eq:glin}.}
\label{fig:modelAFM}
\end{figure*}

\section{Theoretical modeling}

The analysis of force-indentation $F(\delta)$ curves are usually performed wtihin the framework of the Hertz contact theory \cite{hertz,sneddon1965}:

\begin{equation}
\label{eq:Hertz}
F_H(\delta)= \Gamma E^* \delta^\lambda.
\end{equation}

\noindent The subscript H stands for Hertz model, $\Gamma$ and $\lambda$ are geometry-dependent parameters. $E^* = E/(1-\nu^2)$ is the reduced elasticity modulus, and $\nu$ is the Poisson ratio ($\nu=0.5$ for incompressible materials). The elasticity modulus $E$ is related to the shear modulus $G$ as $2 G = E/(1+\nu)$. For pyramidal indenters, one has $\Gamma = 2\tan \theta/\pi$ ($\theta$ being the half-opening angle of the indenter) and $\lambda=2$. For spherical indenters, one has $\Gamma = 4\sqrt{R}/3$ ($R$ being the indenter radius) and $\lambda=3/2$. The schematics of the indentation of soft samples is shown in Figure \ref{fig:modelAFM}.

Hertz theory is based on the following major assumptions: (i) the sample is assumed as a purely elastic half-space, (ii) the stress-strain response is linear, (iii) the elasticity modulus is constant. Therefore, Hertz model is not appropriate to describe viscoelastic materials. Despite of that, several groups have proposed modifications to Hertz model in order to include viscoelastic effects \cite{rebelo2014,darling2006, darling2007, ketene2011, mahaffy2004, alcaraz2002}. For example, Darling \textit{et al.} and Ketene \textit{et al.} studied the viscoelastic properties of living cells using a modified AFM apparatus to perform stress relaxation experiments: they produced an initial indentation in the cells, and recorded the stress relaxation in the cell by monitoring the cantilever deflection as a function of time \cite{darling2006,darling2007,ketene2011}. To explain their measurements, they assumed that the cells could be described by the SLS viscoelastic model, and used the functional method originally proposed by Lee and Radok to obtain the force load history of samples subjected to an instantaneous step indentation (as performed in stress relaxation tests) \cite{leeradok1960}. For this specific case, the force load history essentially keeps the mathematical form of Hertz model, but replacing the fixed value of $E$ by the SLS relaxation function. For other indentation histories, their corresponding force histories must be determined. 

Here we employ the functional method to determine the relaxation properties of soft viscoelastic samples subjected to an indentation history $\delta(t)$ similar to the loading conditions in typical AFM force curves. We derived analytical formulae to extract the viscoelastic parameters directly from the force curves, without any modification in the AFM apparatus to impose a prescribed indentation or load history like in previous reports  \cite{rebelo2014,darling2006, darling2007, ketene2011, mahaffy2004, alcaraz2002}. It is known that the functional method is valid for the cases where the contact area increase monotonically \cite{leeradok1960}. This restriction has been removed by the challenging formulation proposed by Ting \cite{ting1966}. However, Vandamme \textit{et al.} have shown that the functional method not only works well under during the loading phase where the indentation depth $\delta(t)$ is a monotonically increasing function, but it also remains valid remains valid to calculate the initial unloading phase \cite{vandamme2006}. 

In time domain, the elastic Hertz-like model has the form $F(t)=\Gamma E^* \delta^\lambda(t)$. We assume $F(t) = F_{max} \bar{F}(t)$ and $\delta(t) = \delta_{max} \bar{\delta}(t)$, where $F_{max}$ and $\delta_{max}$ represent the maximum load and indentation depth, respectively. $\bar{F}(t)$ and $\bar{\delta}(t)$ represent the load and indentation histories, respectively. In Laplace domain, the associated Hertz-like elastic problem becomes:

\begin{equation}
\label{eq:hertzlaplace}
\tilde{F}(s) =  \frac{\Gamma \delta_{max}^\lambda}{(1-\nu^2)F_{max}} \frac{\tilde{E}(s)}{s} s\tilde{\delta^\lambda}(s).
\end{equation}

\noindent From the constitutive equation of a given viscoelastic model one determines $\bar{E}(s)$, the relaxation function $\tilde{R}(s)=\tilde{E}(s)/s$ and the creep compliance function $\tilde{J}^{-1}(s)=s\tilde{E}(s)$ in Laplace space. Applying the convolution property of the Laplace transform, we obtain:

\begin{equation}
\label{eq:functionaleqgeneral1}
\bar{F}(t) =  \frac{F(t)}{F_0} =  E_0 \int_0^t \bar{R}(t-t^\prime) \frac{d \bar{\delta}^{\lambda}(t^\prime)}{d t^\prime} d t^\prime, 
\end{equation}

\noindent where 
$F_0 = \Gamma \delta_{max}^\lambda/(1-\nu^2)$ is a force normalisation factor, and  $\bar{R}(t) = R(t)/E_0$ with $E_0$ being the instantaneous elasticity modulus. An alternative formulation of Eq.~\ref{eq:functionaleqgeneral1} is obtained by providing a force history  $F(t)$ to determine $\delta(t)$: 

\begin{equation}
\label{eq:functionaleqgeneral2}
\delta^\lambda(t) =  \frac{(1-\nu^2)}{\Gamma}   \int_0^t J(t-t^\prime) \frac{d F(t^\prime)}{d t^\prime} d t^\prime, 
\end{equation}

\noindent where $J(t)$ is the creep compliance function. This approach was adopted by Chyasnavichyus \textit{et al.}, who assumed a constant rate force ramp and the creep compliance function of the SLS model to measure the dynamic properties of PnBMA polymers \cite{tsukruk2014}.

The choice between formulations depends on the working principle of the indenter. Regular nanoindenters apply the load vertically in the same axis of the indenter, aiding the precise control on the load history $F(t)$. In this case, the formulation given by Eq. \ref{eq:functionaleqgeneral2} is more appropriate. In a typical AFM force curve (see Fig.~\ref{fig:modelAFM}), the load is applied indirectly by extending the piezo, and making contact between the cantilever and the sample. The cantilever deflects upward as the piezo extends until a maximum deflection (force) is achieved. Then, the piezo retracts ending the contact between the cantilever and sample. One has precise control in the rate of expansion/retraction of the piezo, but neither in the force nor the indentation histories. 

The indentation depth in AFM force curves are computed as $\delta(t) = [z(t)-z_0] - [d(t)-d_0]$, where $z$ is the piezo displacement, $d$ is the cantilever deflection, and $(z_0,d_0)$ represents the contact point in the force curve. If the sample is infinitely hard, no indentation occur, and the rate of the piezo extension is equal to the rate of cantilever deflection $\dot{z}(t) = \dot{d}(t)$. If the sample is infinitely soft, there is no cantilever deflection, and the indentation rate becomes equal to rate of piezo extension $\dot{z}(t) = \dot{\delta}(t)$. The intermediate cases comprehend  soft samples in general, for which we have $\dot{\delta}(t) = \dot{z}(t)- \dot{d}(t)$. It is known that $\dot{z}(t) \approx \pm 2 L_z f_z$ (for load and unload phases, respectively), where $f_z$ is the vertical scan rate, and $L_z$ is the amplitude of the piezo extension.

Predicting the indentation history that samples undergo during a force curve measurement is difficult. As a first approximation, one can assume a linear indentation history $\delta(t) \propto z(t) =  v t$. The deviation from this behavior can be modeled by a second order contribution, which has small effects on the overall measurement (See Supplementary Material). The linear indentation history approximation leads to simple analytical formulae, which can be easily incorporated to current software packages for the analysis of AFM force curves. One important aspect of indentation experiments with AFM force curves is the detection of the contact point \cite{rudoy2010,benitez2013,roy2014}. Here, we have employed the bi-domain polynomial (BDP) Method of Roy \textit{et al.} which allows a quick method to detect the contact point in the force curves \cite{roy2014}.

\subsection{Viscoelastic modeling}

We adopted the fractional SLS viscoelastic model that comprises either a single relaxation time or a power law relaxation (see Figure \ref{fig:modelAFM}). The choice between relaxation types depends upon a single parameter $0\leq n \leq 1$. The shear relaxation function of the fractional SLS viscoelastic model is given by (See Supplementary Material):

\begin{equation}
\label{eq:SLSrelax}
R(t) = E_0\left[(1 -\alpha) + \alpha E_{n,1}\left[-\left(\frac{t}{\tau}\right)^n\right]\right]. 
\end{equation}

\noindent The shear modulus relaxes from the instantaneous $G(0) =G_0$ to the relaxed modulus $G(t \rightarrow \infty) =G_\infty$, where the amplitude of relaxation is $G_1 = G_0-G_\infty$. $E_{n,1}(z)$ is the generalized Mittag-Leffler function. For $n=1$ one has $E_{1,1}(z)=\exp (z)$ which results in the shear relaxation function of the conventional SLS model  \cite{schiessel1995,shukla2007,jaishankar2012}. $\alpha$ is defined such that $G_1 = \alpha G_0$ and $G_\infty =(1-\alpha) G_0$. It can be regarded as a parameter that describes the \emph{viscoelastic relaxation amplitude} of the material, whereas $\alpha = 0$ represents the elastic limit, and $\alpha = 1$ represents the viscoelastic limit corresponding to the Kelvin-Voigt model. Finally, $\tau$ is a relaxation time. The complex shear modulus is given by $G^*(\omega)= i\omega\int_0^\infty R(t) \exp (-i \omega t ) dt$, and the storage and loss moduli are, respectively:

\begin{eqnarray}
\label{eq:glin}
G'(\omega) =  Re[G^*(\omega)]=G_0\left[ (1-\alpha) + \alpha \frac{ (\omega\tau)^{2n}}{1+(\omega\tau)^{2n}}  \right] \\
G''(\omega)=  Im[G^*(\omega)]=G_0\alpha\left[ \frac{ (\omega \tau)^n}{1+(\omega\tau)^{2n}}  \right]. \nonumber
\end{eqnarray}

\noindent  The storage modulus $G'(\omega)$ suggests that the instantaneous elasticity modulus measured by an AFM force curve should exhibit a frequency-dependent behavior of the form $E_0(\omega) = E_0(1 - \alpha) + E_0\alpha f(\omega)$, such that $f(\omega) \rightarrow 0$ for  $\omega \rightarrow 0$ (in the the limit of very slow piezo extension),  and $f(\omega ) \rightarrow 1$ for $\omega\tau >> 1$ (in the limit very fast piezo extension). There are two crossovers between $G'(\omega)$ and $G''(\omega)$. For $n=1$ and $\alpha \ge 0.82$, they are located at $\omega_{1,2} =(\alpha/ 2 \tau) [1\pm \sqrt{1 - 4(1-\alpha)/\alpha^2}]$.  Figures \ref{fig:modelAFM}(d)-(e) show the general behavior of $R(t)$ and $G^*(\omega)$ for different values of $n$. The exponent $n$ governs the dynamics of $G''(\omega)$, and $\tau = \omega_0^{-1}$ is the inverse frequency for which  $G''(\omega)$ reaches its maximum value. The dynamic viscosity is determined by 

\begin{equation}
\label{eq:eta}
\eta= \frac{G''(\omega)}{\omega} = G_0 \alpha\frac{ \omega^{n-1}\tau^n}{1+(\omega\tau)^{2n}}.
\end{equation}

\begin{figure*}[ht]
\center \includegraphics[scale=0.6]{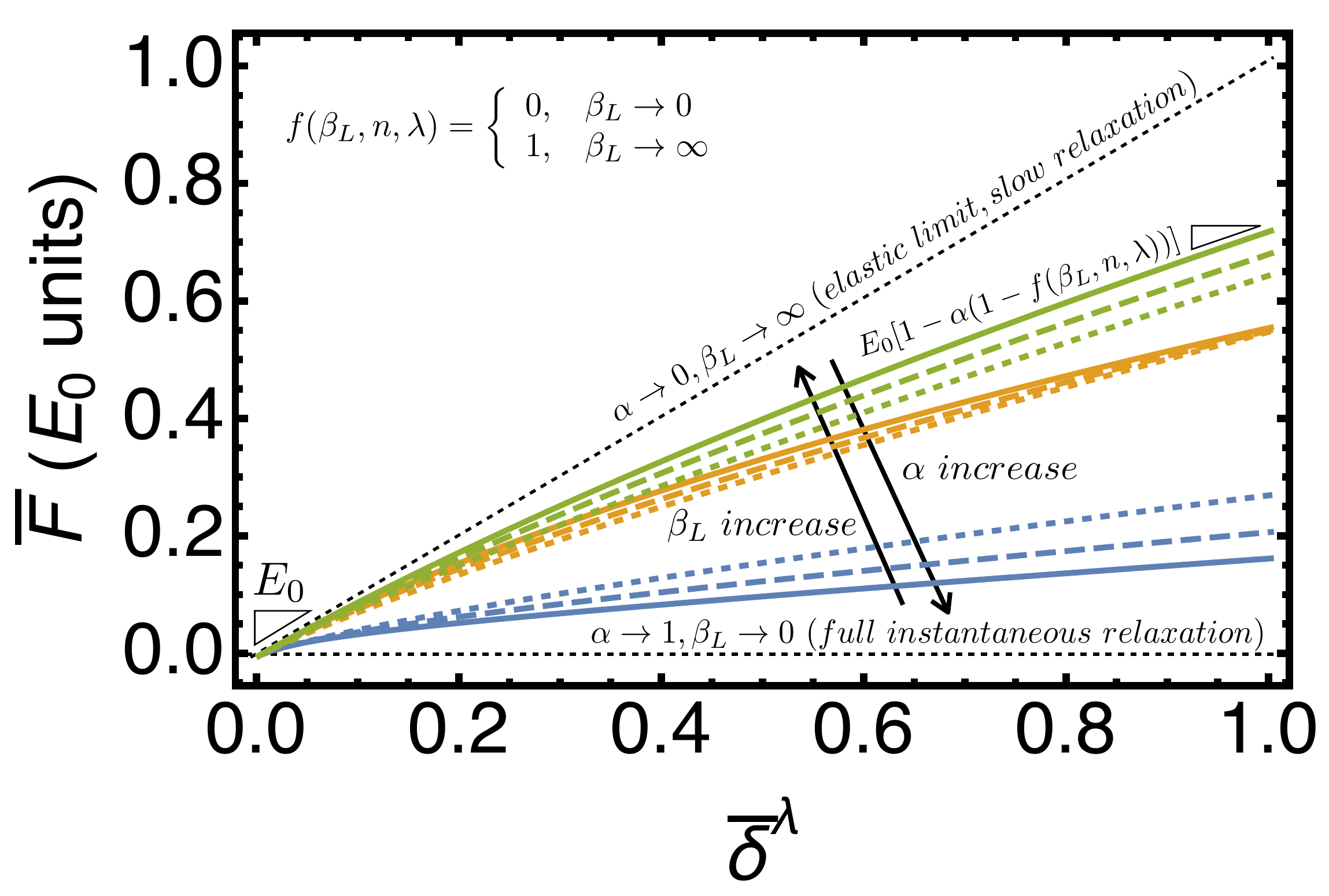}
\caption{Generic force curves exhibiting the role of the parameters $E_0$, $\alpha$, $\beta_L$ and $n$ in the loading  curves. The curves were generated using $\alpha = 0.9$, $\beta_L= 0.05$ (blue), $0.5$ (orange), $1.0$ (green), and $n = 1$ (solid), $0.8$ (dashed), $0.6$ (dotted).}
\label{fig:diagramforce}
\end{figure*}

\subsection{Force curve model}

The linear indentation history during a force curve measurement is given by:

\begin{eqnarray}
\label{eq:deltahistory}
\bar{\delta}_L(t) =  t/\tau_L ~~~(t\leq \tau_L) \\ \nonumber
\bar{\delta}_U(t) = 1 - (t-\tau_L)/\tau_U ~~~(\tau_L \leq t \leq \tau_L+\tau_U),
\end{eqnarray}

\noindent where $\tau_L$ and $\tau_U$ represent the duration of the load (approach) and unload (retract) phases of the force curve.  Replacing $\bar{\delta}_{L}(t)$ in Eq. \ref{eq:functionaleqgeneral1},  the resulting integral can be solved analytically, and the load curves of conical and spherical indenters can be cast in the following expression:

\begin{equation}
\label{eq:forceconeapp}
\frac{\bar{F}_L(\bar{\delta})}{E_0} = \bar{\delta}^\lambda\left[  1 - \alpha  +\alpha \Gamma(\lambda+1) E_{n,\lambda+1}\left[-\left(\frac{\bar{\delta}}{\beta_L}\right)^n   \right]     \right],
\end{equation}

\noindent where $\Gamma(z)$ is the Gamma function, and $E_{n,m}(z)$ is the generalised Mitta-Leffler function \cite{schiessel1995,shukla2007,jaishankar2012}. We simplified notation by making $\bar{\delta}_{L}(t) \rightarrow \bar{\delta}$, and $\beta_{L}=\tau/\tau_{L}$.  Although the validity of the functional method for the unload curve is not completely understood except near $\bar{\delta}=1$ \cite{vandamme2006}, we provide analytical formulas for the unload curves for the specific case of $n=1$ in the Supplementary Material.

%
%
%
%


The force curves $d~versus~z$ must be transformed into the form $\bar{F}~versus~\bar{\delta}^\lambda$, where $\bar{F} = k_c(d-d_0)/F_0$, $F_0 = \Gamma \delta_{max}^\lambda/(1-\nu^2)$, and $\bar{\delta}=\delta/\delta_{max}$. We must also compute $\tau_L$ directly from the force curves. This results in universal curves as the ones shown in Figure \ref{fig:diagramforce}. In the elastic limit (by elastic we mean either a truly elastic material or a viscoelastic material whose relaxation time is much longer than $\tau_L$), the force curve is a straight line whose slope is $E_0$. The slope of the force curve is $E_0$ for $\bar{\delta} \rightarrow 0$ and $E_0[1-\alpha(1-f(\beta_L,n,\lambda))]$  for $\bar{\delta} \rightarrow 1$. The function $f(\beta_L,n,\lambda) = \Gamma (\lambda+1) E_{n,\lambda+1}(-\beta_L^{-n})$ has a weak dependence on the indenter geometry and exhibits the following behavior: (i) $f(\beta_L,n,\lambda)\rightarrow 0$ for $\beta_L\rightarrow 0$, and (ii) $f(\beta_L,n,\lambda)\rightarrow 1$ for $\beta_L\rightarrow \infty$. At the opposite end of the elastic limit, one has the full instantaneous relaxation case obtained when $\alpha \rightarrow 1$ and $\beta_L \rightarrow 0$. For fixed values of $\alpha$ and $n$, increasing values of $\beta_L$ make the curves to move towards the elastic limit case. For a fixed values of $\beta_L$ and $n$, increasing values of $\alpha$ make the curves to move towards the full instantaneous relaxation case. We remark that the aforementioned behavior are general trends of force curves taken in fractional SLS viscoelastic materials (see Figure \ref{fig:diagramforce}), and are valid even for nonlinear indentation profiles.

\begin{figure*}[ht]
\center \includegraphics[scale=0.70]{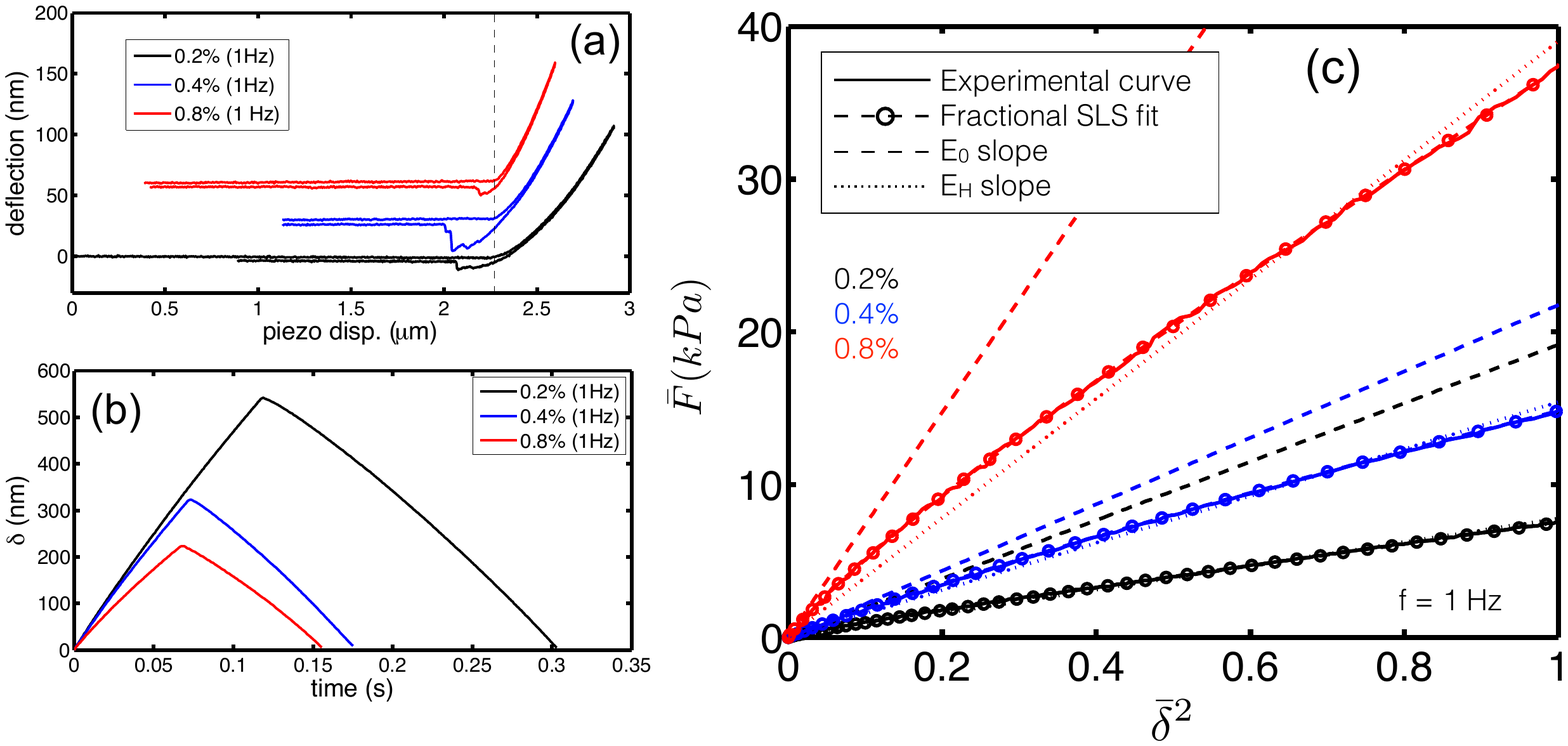}
\caption{(a) Comparison of AFM force curves measured in polyacrylamide gels with different $C_{bis}$. The vertical scan frequency is $f_z = 1$Hz. A maximum deflection of $d_{trigger}=100$ nm was imposed in all measurements. From those curves, we constructed (b) their respective indentation histories $\delta(t)$.
(c) Normalised force curves of polyacrylamide gels. Solide lines represent experimental curves, dashed lines represent the fitting with Hertz model, and symbols represent the fitting with different relaxation exponents. The parameters extracted from those curves are shown in Table \ref{tab:poliac1hz}.}
\label{fig:poliac1hz}
\end{figure*}

\begin{table*}[ht]
\caption{List of viscoelastic parameters extracted from the force curves of Figure \ref{fig:poliac1hz}(c). The lines lacking data of the parameters $\alpha$,  $\tau$, $\beta_L$ and $n$ represent the data obtained with the conventional Hertz fit.}
\begin{center}
 \begin{tabular}{@{} crrrrrrrrr @{}}
    \hline
    \hline
 $C_{bis}$  & $f_z$( Hz) &  $E_{0}$ (kPa) & $\alpha$ & $\tau$(s)& $\beta_L$ & $r^2$ & RMSE & $n$ \\
    \hline

0.2\% & 1 &    7.778  &  &  &  & 0.9952   & 0.1575 & \\
0.4\% & 1 &    15.402  &  &  &  & 0.9960 & 0.2880 & \\
0.8\% & 1 &    39.404  &  &  &  & 0.9921 & 1.0189 & \\ \hline

0.2\% & 1 &    19.154  & 0.677  & 0.005 & 0.043 & 0.9997 & 0.0393 & 0.91\\
0.4\% & 1 &    21.751  & 0.931  & 0.072 & 0.863 & 0.9998 & 0.0681 & 0.79\\
0.8\% & 1 &     73.645  & 0.638 & 0.009 & 0.129 & 0.9999 & 0.1375 & 1\\

\hline
\hline

  \end{tabular}
\end{center}
\label{tab:poliac1hz}
\end{table*}%

\section{Experimental validation}

\subsection{Experimental details}

Polyacrylamide gels were prepared from the stock solution of 30\% acrylamide (by weight) in three different concentrations of N,N-bis acrylamide (0.2\%, 0.4\% and 0.8\% in volume). The final gels have a concentration of 0.375 M Tris-HCL, with pH 8.8. The gels were polymerized chemically by addition of 10$\mu$L of tetramethylethylenediamine (Temed) and 0.1 mL of 15\% ammonium persulfate solution/10mL of gel solution.

An AFM (MFP-3D, Asylum-Research, Santa Barbara, CA, USA) was used to measure conventional force curves. Soft cantilevers (Microlever, MLCT-AUHW, Veeco, USA) with a spring constant of 0.015 N/m were used to probe the gels, verified by the thermal method \cite{radmacher2007}. A maximum force trigger of $d_{trigger}=100$nm were imposed to avoid excessive gel indentation. The used AFM tips have pyramidal shapes with half-opening angles of $\theta= 38^\circ$. To reduce adhesion effects in the cantilever, the force measurements were performed in distilled water. Different vertical scan frequencies (0.1Hz, 0.5Hz, 1Hz, 5Hz and 10Hz) were applied to samples to investigate the effect of varying load rates in the viscoelastic properties. The frequency-dependent data were averaged over different ($n=5$) locations for each gel.

\begin{figure*}[ht]
\center \includegraphics[scale=0.6]{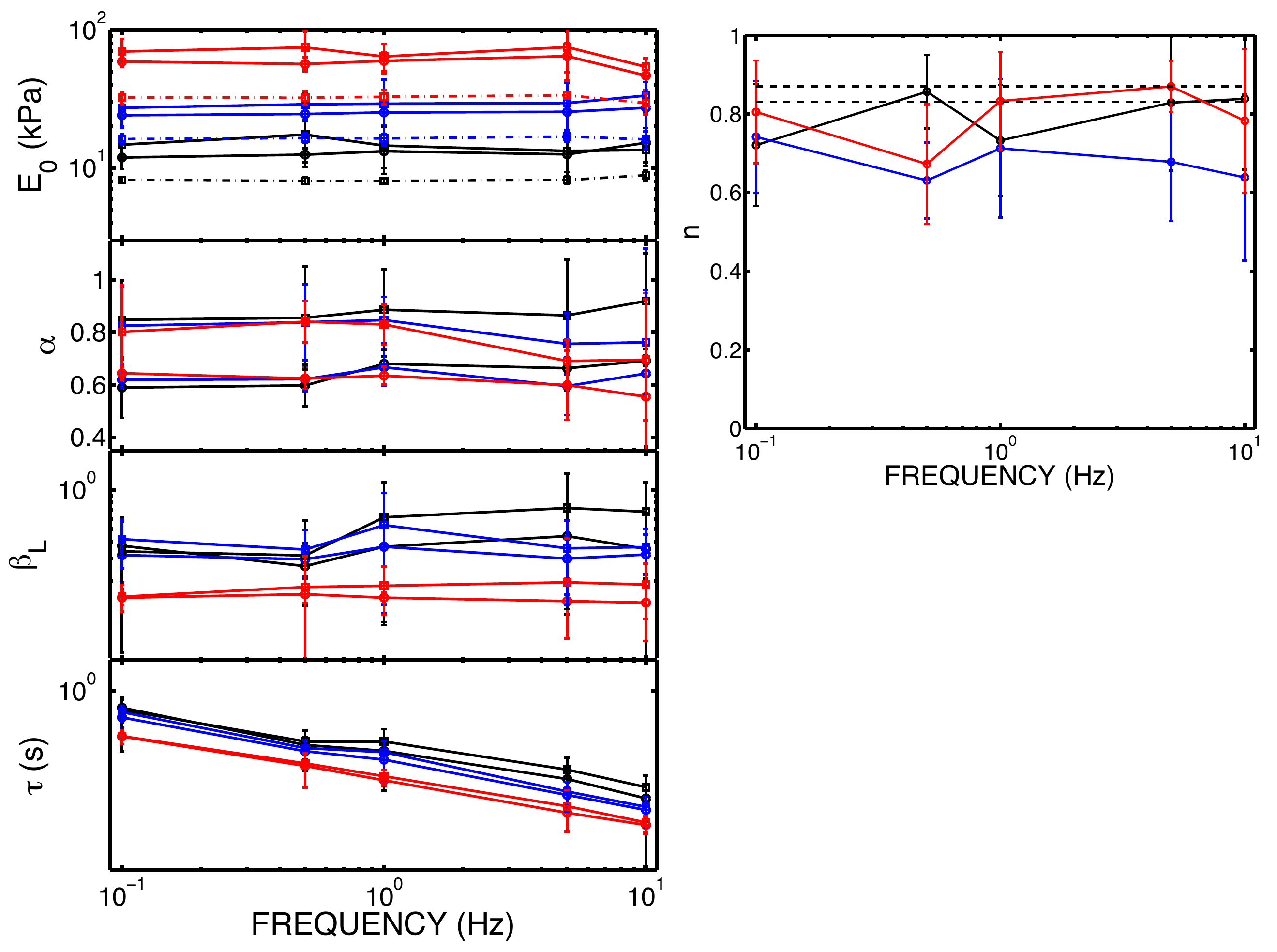}
\caption{(left) Frequency dependence of $E_0$, $\alpha$, $\beta_L$ and $\tau$. Square (circle) symbols represent the fitting with fractional (regular) SLS viscoelastic model. Colors represent different $C_{bis}$ concentrations: 0.2\% (black), 0.4\% (blue), 0.8\% (red). The dash-dotted lines in panel (a) represent the fitting go $E_0$ with Hertz model. (right) Frequency dependence of the relaxation exponents. The horizontal lines represent the exponents (0.83 and 0.87) extracted from the polyacrylamide gels measured by Abidine \textit{et al.} \cite{abidine2015} (See Table \ref{tab:abidine}).}
\label{fig:rheology}
\end{figure*}

\subsection{Study of polyacrylamide gels}

Figure \ref{fig:poliac1hz}(a) shows AFM force curves (with their respective contact points aligned) of polyacrylamide gels measured with $f_z=1$Hz, and their respective time-dependent indentation profiles $\delta(t)$ are shown in Figure \ref{fig:poliac1hz}(b). The comparison of the indentation depths clearly shows that the increase of the bisacrylamide concentration ($C_{bis}$) enhances the stiffness of the gels. The indentation profile is not linear, and the time to reach the maximum indentation are also different, suggesting distinct viscoelastic relaxation properties among samples. We tested the validity of the linear indentation approximation in all measurements in this work. We obtained that all force curves exhibited very small deviation from linear behavior, lying within a quasi-linear indentation regime. Therefore, we can use Eq. \ref{eq:forceconeapp} to extract the viscoelastic parameters of the polyacrylamide gels at the expense of very small errors. The error analysis of the linear indentation approximation is discussed in the Supplementary Material. 

Figure \ref{fig:poliac1hz}(c) shows the $\bar{F}~versus~\bar{\delta}^2$ curves for different $C_{bis}$ measured with $f_z = 1$ Hz.  The comparison of $E_0$ slopes shows that the increase of $C_{bis}$ enhances the gel stiffness. The relaxation times can be qualitatively compared by how close to $\bar{\delta}=0$ the force curve deviates from the $E_0$ slope. Therefore, the largest relaxation time must be observed in the 0.4\% sample, while samples 0.2 and 0.8\% should exhibit comparable values of $\tau$. The same trend is observed in the value of $\beta_L$.  All viscoelastic parameters fitted from Figure \ref{fig:poliac1hz}(c) are listed in Table \ref{tab:poliac1hz}. For these individual force curves, we obtained exponents $n$ varying between 0.79 and 1.0.

The dynamic behavior of the gels is shown in Figure \ref{fig:rheology}. Here we adopted two fitting strategies. First, we assumed that gels can be described by the conventional SLS model (exponent $n=1$). Second, we assumed that the gels can be described by the fractional SLS model where $n$ is also a fitting parameter. In both cases, the stiffness of the gels are proportional to $C_{bis}$, and do not exhibit any appreciable frequency dependence. The Hertz fitting of the force curves provided lower values  $E_H < E_0$ compared to our model. This is due to viscoelastic relaxation that reduces the instantaneous elasticity modulus during the course of the force curve measurement. This can be better seen in Figure \ref{fig:poliac1hz}(c) that shows that the slope of the force curves near $\bar{\delta}\rightarrow 0$ is always larger than the slope of the Hertz fit. In the limit of very slow piezo extension, one obtains $E_H \approx E_\infty$. For very fast piezo extension, Hertz fitting provides $E_H \approx E_0$. The nearly constant difference $E_0-E_H$ suggests that the relaxation times of the gels are much shorter than 0.1 s (only accessible for frequencies above 10 Hz). This figure also shows that The relaxation times are inversely proportional to $C_{bis}$, exhibiting power law decay, while $\alpha$ and $\beta_L$ are nearly frequency independent.

The comparison of both fitting strategies show that $E_0$, $\alpha$ and $\tau$ are inversely proportional no the exponent $n$. This is reasonably simple to understand with the help of the relaxation functions in Figure \ref{fig:modelAFM}(d). Although similar, those curves are not exponential decaying functions, except for the case $n=1$. The determination of $E_0$ is governed by the derivative of the force curve $dF_L/d\bar{\delta^\lambda}$ at $t \rightarrow 0$, which is roughly proportional to $R_{n}(t)$ (note that $R(t)$ depends $E_{n,1}[-(t/\tau)^n]$, while $F_L(t)$ depends on $E_{n,\lambda+1}[-(t/\tau)^n]$, but the shape of those function are qualitatively similar to each other). An attempt to fit the relaxation function $R_{n<1}(t)$ with an exponential decay leads to overestimated values near $t = 0$ in order to reduce the total fitting error, resulting in an underestimated value of $E_0$. The functions $R_{n<1}(t)$ seem to converge to values $R_{n<1}(t\rightarrow\infty) > 1 - \alpha_{n<1}$, this will lead to underestimated values of $\alpha$. The overall adjustment of the curves $R_{n<1}(t)$ with a exponential decay demands an steeper descend behavior near $t=0$, that leads to underestimated values of $\tau$.

The fitted relaxation exponents (for the cases where $n$ was also a fitting parameter) are shown in Figure \ref{fig:rheology}.  Despite of the fluctuations, the relaxation exponents are nearly frequency independent between 0.1 Hz and 10 Hz. The average exponents for the 0.2\%, 0.4\% and 0.8\% samples are 0.796, 0.681 and 0.793, respectively. The average exponent for all $C_{bis}$ cases is 0.756. The fluctuations and error bars are attributed to spatial non-homogeneities in the gel composition, and other complicated effects which are not included in our model like hydrodynamic interaction of the cantilever with liquid (all measurements were performed in liquid to minimize the adhesiveness of the gels) and residual adhesive effects.

\section{Discussion}

Abidine \textit{et al.} investigated the dynamic rheology of polyacrylamide gels in a wide range of frequencies combining conventional shear rheometry (for low frequencies ranging between 10$^{-3}$ Hz and 1 Hz), and AFM based dynamic indentation (for high frequencies ranging from 1 Hz to 300 Hz) \cite{abidine2015}. They obtained elasticity moduli varying from 1 kPa to 30 kPa for bisacrylamide concentrations (in weight) ranging between 5\% and 15\%. They have shown that the storage modulus exhibits a constant plateau between 10$^{-3}$ Hz and 50 Hz, for all bisacrylamide concentrations. The storage shear modulus $G'(\omega)$ is one order of magnitude higher than the loss modulus $G''(\omega)$ from frequencies up to 100 Hz, above which there is a crossover between $G'(\omega)$ and $G''(\omega)$. It is instructive to fit their rheological measurements with the fractional SLS relaxation model in order to extract quantities that can be compared to ours. 

\begin{table*}[ht]
\caption{SLS parameters extracted from the dynamic measurements of the gels measured by Abidine \textit{et al.} \cite{abidine2015}.}
\begin{center}
 \begin{tabular}{crrrrrrrr}
    \hline
    \hline
  $C_{bis}$     & $E_{0}$(kPa) & $E_{\infty}$(kPa) & $E_{1}$(kPa) & $\alpha$ & $\tau$ (s) & $r^2_{G'}$  & $r^2_{G''}$ & $n$\\
    \hline
   \multicolumn{9}{c}{conventional SLS model ($n = 1$)} \\
    \hline
 5\%   &    15.816  & 1.290    & 14.525  & 0.918 & 3.66 $\times 10^{-4}$  & 0.9945   & 0.9970 & 1\\
7.5\% &    19.996  & 2.086    & 17.910  & 0.895 & 5.48 $\times 10^{-4}$  & 0.9974   & 0.9972 & 1\\
   \hline
   \multicolumn{9}{c}{fractional SLS model ($n$ is a fitting parameter)} \\
    \hline
 5\%   &    28.056  & 1.274    & 26.781    & 0.954 & 1.36 $\times 10^{-4}$ & 0.9963  & 0.9998 & 0.87 \\
7.5\% &    29.734  & 2.079    & 27.654    & 0.930 & 2.29 $\times 10^{-4}$ & 0.9984  & 0.9994 & 0.83 \\
    \hline
    \hline
  \end{tabular}
\end{center}
\label{tab:abidine}
\end{table*}%

\begin{figure*}[ht]
\center \includegraphics[scale=0.6]{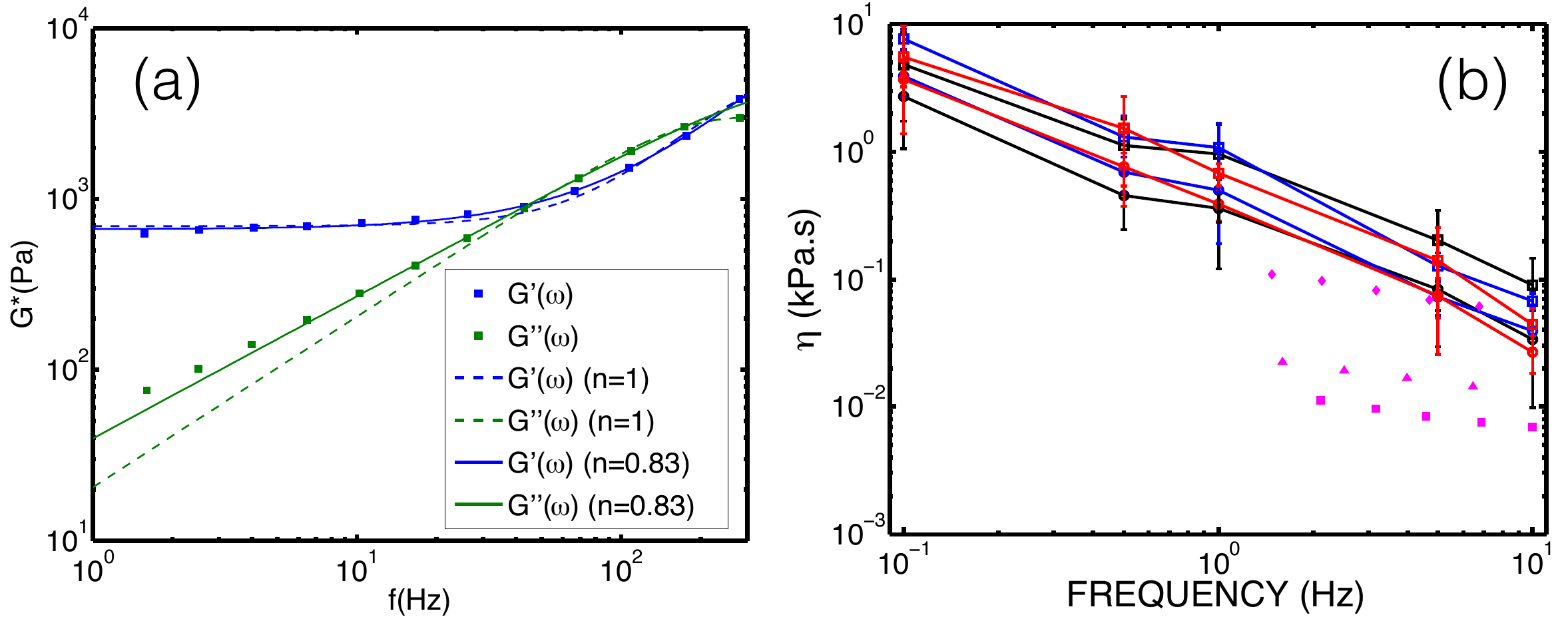}
\caption{(a) AFM based dynamic rheology of polyacrylamide gels (7.5\% of bisacrylamide) measured by Abidine \textit{et al} \cite{abidine2015}. Solid lines represent the fitting of the data with the complex shear modulus of the SLS relaxation function (power law relaxation exponent $n=1$). Dashed lines represent the fitting of the data with complex shear modulus of the fractional-like SLS relaxation function (power law relaxation exponent $n=0.83$). Table \ref{tab:abidine} shows the fitted parameters from Abidine's gels. (b) Frequency dependence of the apparent viscosity $\eta=E_0 \alpha \tau$. Colors represent different $C_{bis}$ values: 0.2\% (black), 0.4\% (blue), 0.8\% (red). Square (circle) symbols represent the fitting with fractional (regular) SLS viscoelastic model. Pink symbols represent the dynamic viscosity of Abidine's gels calculated as $\eta(\omega) = G''(\omega)/\omega$: square (5\%), triangle (7.5\%), diamond (15\%). }
\label{fig:abidine}
\end{figure*}

Figure \ref{fig:abidine} shows an example of the rheological data of Abidine \textit{et al.} Both regular and fractional SLS models are able to describe well the storage modulus in the whole frequency range, with the fractional model exhibiting a slightly improved accuracy. The conventional SLS model fails to describe $G''(\omega)$ for frequencies up to 10 Hz, which is precisely the largest frequency that the forces curves can be measured in our AFM. On the other hand, the fractional model (with $n = 0.83$) describes $G''(\omega)$ very well between 1 Hz and 300 Hz. Abidine's data suggests that an exponent a little smaller than 0.83 between 1 Hz and 10 Hz. The viscoelastic parameters fitted from Abidine's measurements are shown in Table \ref{tab:abidine}. We focused our comparison in Abidine's gels with lowest $C_{bis}$ (5.0\% and 7.5\%) because the cut-off frequency $\omega_0$ is out of the measured frequency range for gels with higher $C_{bis}$. 

The fitted values of $\omega_0^{-1}$  of the order of $10^{-4}$ s confirm that our measurements were performed much below $\omega_0$, in a frequency range in which the instantaneous elasticity modulus must be almost fully relaxed. The $E_0$ values in both works are in good agreement, and in both studies $E_0$ is proportional to $C_{bis}$, while $\alpha$ and $\tau$ are inversely proportional to $C_{bis}$. The fitted values of $\alpha$ are compatible ($\alpha > 0.82$) with the values for which there is a double crossover between $G'(\omega)$ and $G''(\omega)$ (see Figure \ref{fig:abidine}). The parameters $E_0$, $\alpha$ from Abidine's measurements are in good agreement with the values estimated from our measurements. The $\tau$ values obtained estimated from our model are of the order of $10^{-3}$ s, nearly one order of magnitude higher than the $\omega_0^{-1}$ values estimated from Abidine's measurements. One possible reason for this is that the time resolution of the AFM force measurements is limited by the reading frequency of the AFM controller of 2 kHz.

We also obtained a very good agreement between the exponents from the frequency dependent AFM force curves (ranging between 0.681 and 0.796) and the exponents from Abidine's measurements (ranging between 0.83 and 0.87).  We remind that our exponents were fitted from rheological data acquired between 0.1 Hz and 10 Hz, while Abidine's exponents were fitted from data in the range 1 Hz to 300 Hz. However, a quick look in Abidine's $G''(\omega)$ curves suggests that slightly lower exponents would be obtained data between 1 Hz and 10 Hz, becoming even closer to our values.

Finally, Eq. \ref{eq:eta} suggests that the dynamic viscosity can be written in form $\eta(\omega) = G_0(\omega) \alpha(\omega) \tau(\omega)$, where $E_0 = 2(1+\nu)G_0$, $\alpha$ and $\tau$ are the frequency-dependent parameters extracted from the force curves. Figure \ref{fig:abidine}(b) compares our measured effective viscosities with Abidine's data. Both studies exhibit very good qualitative agreement, with similar slopes. The quantitative discrepancy is attributed to different gel composition between both studies.

\section{Conclusions}

We derived an analytical force-indentation model to describe viscoelastic materials with power law relaxation, that can be easily incorporated in the analysis of AFM forces. The major approximation to derive this model is that the indentation profile is linear. This approximation is valid  as long as the degree of nonlinearity of the indentation profile is small. The force model is in excellent agreement with FEM simulations of viscoelastic materials indented by conical and spherical indenters (see Supplementary Material). Experimentally, we tested the model by measuring the viscoelastic properties of polyacrylamide gels with AFM force curves with varying load load frequencies at room temperature. The viscoelastic properties of the gels exhibited very good agreement with the results of Abidine \textit{et al.} \cite{abidine2015}. A very important characteristic of the model is that it was able to reproduce the viscoelastic properties of the gels, regardless the measurement method. For example, we have used simple AFM force curves, while Abidine \textit{et al.} used a dynamic indentation experimental method based on a custom modification of the AFM apparatus to measure $G'(\omega)$ and $G''(\omega)$. Most strikingly, the model is able to determine the viscoelastic relaxation exponent without a direct measurement of $G''(\omega)$. Our exponent values ranging 0.681 and 0.796 (measured between 0.1 Hz and 10 Hz) are in very good agreement with the exponents ranging between 0.83 and 0.87 from Abidine's data (measured between 1 Hz and 300 Hz). 

In principle, force curve based rheology is limited to low loading frequencies (up to 30 Hz) to avoid strong cantilever oscillations during approach and retract motions, but one can use the time-temperature superposition principle by performing force measurements at different temperatures to study the viscoelastic response of the materials in a much wider range of frequencies. This method was recently demonstrated by Chyasnavichyus \textit{et al.} \cite{tsukruk2014}. Finally, the proposed model is simple enough to be easily incorporated in AFM data analysis softwares.

\noindent \textbf{Acknowledgements.}  The authors acknowledge the financial support from the Brazilian National Research Council (CNPq).


\pagebreak

\bibliography{refs}

\end{document}